\documentclass[sigconf]{acmart}
\usepackage{hyperref}
\usepackage{csquotes}
\usepackage{tablefootnote}
\usepackage{booktabs,caption}
\usepackage[flushleft]{threeparttable}
\AtBeginDocument{%
  \providecommand\BibTeX{{%
    \normalfont B\kern-0.5em{\scshape i\kern-0.25em b}\kern-0.8em\TeX}}}

\setcopyright{acmcopyright}
\copyrightyear{2022}
\acmYear{2022}
\acmDOI{XXXXXXX.XXXXXXX}

\acmConference[IEEE/ACM CHASE '22]{IEEE/ACM international conference on Connected Health: Applications, Systems and Engineering Technologies}{November 17--19,
  2022}{Washington, D.C.}
\acmBooktitle{IEEE/ACM international conference on Connected Health: Applications, Systems and Engineering Technologies,
 November 17--19, 2022, Washington, D.C.} 
\acmPrice{15.00}
\acmISBN{978-1-4503-XXXX-X/18/06}

\begin{document}

\title{Fall Detection from Audios with Audio Transformers}

\author{Prabhjot Kaur}
\email{prabhjotkaur@wayne.edu}
\affiliation{%
  \institution{Wayne State University}
  \city{Detroit}
  \state{Michigan}
  \country{USA}
  \postcode{43017-6221}
}

\author{Qifan Wang}
\email{wqfcr@fb.com}
\affiliation{%
  \institution{Meta AI}
  \city{Menlo Park}
  \state{California}
  \country{USA}}

\author{Weisong Shi}
\email{weisong@wayne.edu}
\affiliation{%
  \institution{Wayne State University}
  \city{Detroit}
  \state{Michigan}
  \country{USA}}

\renewcommand{\shortauthors}{Kaur, et al.}

\begin{abstract}
  Fall detection for the elderly is a well-researched problem with several proposed solutions, including wearable and non-wearable techniques. While the existing techniques have excellent detection rates, their adoption by the target population is lacking due to the need for wearing devices and user privacy concerns. Our paper provides a novel, non-wearable, non-intrusive, and scalable solution for fall detection, deployed on an autonomous mobile robot equipped with a microphone. The proposed method uses ambient sound input recorded in people's homes. We specifically target the bathroom environment as it is highly prone to falls and where existing techniques cannot be deployed without jeopardizing user privacy. The present work develops a solution based on a Transformer architecture that takes noisy sound input from bathrooms and classifies it into fall/no-fall class with an accuracy of 0.8673. Further, the proposed approach is extendable to other indoor environments, besides bathrooms and is suitable for deploying in elderly homes, hospitals, and rehabilitation facilities without requiring the user to wear any device or be constantly "watched" by the sensors.
\end{abstract}



\maketitle

\section{Introduction}
A Fall is "an event which results in a person coming to rest inadvertently on the ground or other lower-level" \cite{who2021factsheet}. The current estimates from the World Health Organization (WHO) show that falls are the second major cause of death globally, with approximately 684,000 individuals dying every year. While people of all ages are at risk of falls and related injuries, the population aged 60 and older suffers the highest death rate due to a fall event. Additionally, non-fatal falls are equally devastating as they may lead to a permanent disability, psychological fear of falling, and hospitalization among individuals who have experienced a fall previously. About 37.3 million falls require medical attention each year, and 0.65 million result in deaths globally \cite{who2021factsheet}. Such consequences of falls make them a significant public health problem that needs attention. While falls are not always preventable, their negative impact can be minimized with a prompt response, including detecting a fall event and notifying health support services.

There is a keen interest from both academia and industry in developing techniques for fall detection amongst the elderly \cite{wang2020elderly}. The goal is to detect a fall within a reasonable period after its occurrence so that mitigation steps can be taken to make the consequences of a fall less severe. Such techniques rely on different sensors, broadly categorized into wearable and non-wearable sensors. The user must wear wearable sensors to facilitate the collection of accelerometer/gyroscope data for further analysis. The non-wearable sensors, on the other hand, are external sensors, including vision-based, floor-based, and acoustic sensors. The external sensors monitor the surroundings in their vicinity to detect if a fall event has occurred. Besides the types of sensors used, many different algorithms and methods exist to process the collected data to classify if the sensor input represents a fall. These methods range from traditional statistical methods to more advanced machine learning \cite{usmani2021latest} and deep learning \cite{islam2020deep}, \cite{ribeiro2022iot}, \cite{lafuente2022rgb} based methods. The wearable and vision-based techniques have certain limitations. For example, it is not always feasible to collect data with a wearable recording device as the user may forget to put it on or may need to take it off for performing some activity. Similarly, the vision-based sensors are not scalable as the subject has to be in the sensor's line of sight to be detected, and they invade the user's privacy. These limitations are even more prominent in the bathroom environments, where falls amongst elderly are the most frequent \cite{moles2020characteristics}, \cite{shaharudin2018falls}, \cite{muhammad2020prevalence}, \cite{moreland2021descriptive}. Thus, a fall detection system for the elderly that is accurate, scalable, and preserves user privacy remains an open research problem despite the number of sensors and algorithms developed to date. This paper proposes a non-wearable and scalable solution for fall detection in bathroom environments, implemented on an autonomous mobile robot. The core of the proposed methodology is an ambient sound input that is analyzed to detect fall events. The use of sound input for fall detection offers an advantage over wearable and vision-based sensors. It does not require wearing a recording device, is less invasive than the vision-based methods, and does not require the subject to be in sight to detect falls. Further, the proposed solution is highly scalable as it is implemented on an autonomous mobile robot, thus allowing us to detect a wider range of locations in indoor environments.

\subsection{Problem statement}
The overall objective is to detect falls in bathroom environments that are highly prone to fall events among the elderly. We formulate the problem as an audio classification problem consisting of two classes, namely, fall (positive) and no-fall (negative) classes. The goal is to develop a method that takes audio from the target environment as input and correctly classifies the input into one of the two classes. If a fall event is detected, an alert must be sent to the registered party for their attention.

\subsection{Motivation and Contribution}
Our motivation is to fill the gap in the existing research within the fall detection domain, as the current techniques have limitations regarding their deployment efficacy in bathroom environments. To overcome the need to wear a device and respect user privacy, we use sound input to solve the audio classification problem. Specifically, we develop and train a deep learning model based on the Transformer \cite{vaswani2017attention} architecture that learns to identify fall events from the ambient sound input. The trained model is deployed on an autonomous mobile robot capable of navigating to the target (bathroom) environment and sending a phone text to an appropriate party if help is needed. The main contributions of the paper are as follows:
\begin{itemize}
    \item A non-wearable, non-invasive, and scalable method for detecting fall events in bathroom environments implemented on an autonomous mobile robot. 
    \item A novel audio classification Transformer model that takes noisy environmental sounds as input and correctly classifies them into fall and no-fall classes with an accuracy of 0.8673.
    \item A new set of features for audio classification, which we refer to as ``Diff'' features, that show better performance than the traditional features such as log mel spectrograms and raw audio features commonly used for audio classification \cite{droghini2018few}.
\end{itemize}

It is important to note that while the developed method is demonstrated for bathroom environments, it is extendable to all indoor environments with appropriate training data.

The remainder of the paper is structured as follows. Section \ref{relatedWorks} summarizes the related works. Section \ref{dataset} provides details on the audio dataset used in our work, followed by Section \ref{methodology} that describes the methodology developed to detect fall events using the audio input. Section \ref{implementation} focuses on implementation details of the proposed method in the real world. Section \ref{results} dives into the results of the method developed in this paper. Finally, Section \ref{conclusion} concludes the work.

\section{Related Works}
\label{relatedWorks}
Research in fall detection has gained much momentum in the past few decades. Many researchers have proposed several solutions to detect falls among the elderly. This section briefly reviews various sensors and algorithms used in this space. We emphasize on the more recent works and direct the reader  to existing literature surveys \cite{ren2019research}, \cite{el2013fall}, \cite{chen2018evaluating}, \cite{wang2020elderly}, \cite{zhang2013honey} for a more depth historical advancements in this domain.
\subsection{Fall detection sensors and algorithms}
The techniques for fall detection are classified into two categories based on the underlying sensor technology, namely wearable and non-wearable (external) \cite{delahoz2014survey}. The wearable techniques monitor human movement by constantly collecting accelerometer and gyroscope data from devices the user wears. Such techniques commonly use threshold-based methods to distinguish falls from the normal movement of a human subject. Various literature surveys such as  \cite{karar2022survey}, \cite{delahoz2014survey}, \cite{gray2021development} provide a thorough research in this area. One of the major advantages the wearable sensors offer is that they are portable, making them highly scalable as long as the user wears the device. Additionally, wearable sensors are cost-effective. It is established that wearable technology performs quite well, gaining more than 90\% accuracy  for detecting falls  \cite{de2022bluetooth}, \cite{pillai2022wearable}, \cite{miawarni2022towards}, \cite{wu2022applying}. However, the natural drawback of such techniques is that they are intrusive, as the user has to wear them at all times to facilitate data collection. There is also a risk of forgetting to wear them, defeating the purpose of constant monitoring. The non-wearable techniques, on the other hand, overcome these challenges as they are non-intrusive, and there is no concern of forgetfulness by the user. The research in non-wearable technology is equally rich as wearable technology.

The non-wearable sensors for fall detection include wide options, such as vision/camera-based sensors, including multi-cameras and depth cameras, that constantly stream video of the surroundings to identify falls \cite{zhang2015survey}, \cite{ezatzadeh2017fall}. The more recent vision-based solutions typically formulate fall detection as a human pose estimation and recognition problem as in \cite{fei2022flow}, \cite{lafuente2022rgb}. \cite{lafuente2022rgb} proposes a system implemented on a mobile robot that uses its onboard RGB camera to identify falls. The underlying method is based on the You Only Look Once (YOLOv3) deep learning model trained on the extended Fall Person Dataset (E-FPDS). The system achieves a recall of 98.7\% on the test set. Similarly, \cite{fei2022flow} proposes a Flow-pose Net model which combines optical flow and human pose information to achieve more robust detection. The experiments show that the system achieves nearly 100\% accuracy on two public datasets. The cons of using vision-based sensors, however, are that they are sensitive to illumination and occlusion, require synchronization for multi-cameras and invade user's privacy \cite{ezatzadeh2017fall}. To overcome such limitations, researchers have also explored acoustic sensors for indoor monitoring and fall detection \cite{8891779}, \cite{vuegen2019contactless}, \cite{alex2021deep}. Unlike vision-based sensors, these sensors do not suffer from illumination and occlusion problems. Finally, there is a recent trend that uses Internet of Things (IoT) based solutions, which aim to overcome the shortcomings of both wearable and vision-based solutions \cite{karar2022survey}.The most recent IoT-based solution is proposed in \cite{ribeiro2022iot}. The custom-made IoT device measures floor vibrations and sound level where it is installed to detect falls. The method achieves 92.5\% accuracy and no false negatives using Morlet wavelet and 79.01\% accuracy with some false negatives using an Artificial Neural Network (ANN). All of the techniques mentioned above require the sensors to be installed in multiple places for wider coverage. The use of a mobile platform with its onboard sensors, such as in \cite{lafuente2022rgb} provides a more efficient solution.

Besides the sensor technology used for fall detection, multiple algorithmic approaches exist to process sensor data to identify falls. The traditional approaches use statistical methods. While these are easier to implement, they are susceptible to noisy data and generate more false alarms \cite{usmani2021latest}. These approaches are also less efficient in nonlinear problems. The more recent approaches include Machine Learning \cite{usmani2021latest} and Deep Learning-based methods \cite{islam2020deep},  \cite{lafuente2022rgb}, \cite{fei2022flow}, \cite{miawarni2022towards}, \cite{wu2022applying}.

\subsection{Fall datasets}
Access to fall datasets is an ongoing challenge in this domain, making it difficult to compare different methods. Most researchers end up collecting their own dataset to test and evaluate the proposed methods \cite{karar2022survey},  \cite{de2022bluetooth}, \cite{casilari2017umafall}. However, there are few datasets for fall detection that are publicly available. In the case of wearable sensors, SisFall \cite{sucerquia2017sisfall} is one of the largest datasets. It consists of data collected from two accelerometers and a gyroscope, fixed as one compact device to a belt on the participant's waist. It includes 15 types of falls and 19 types of Activities of Daily Living (ADL) collected from 38 volunteers. Other notable datasets from wearable devices are Mobifall, tFall, DLR \cite{casilari2017analysis}, UMAFall \cite{casilari2017umafall} datasets . Similarly, \cite{lafuente2022rgb} releases an extended vision-based Fall Person Dataset (E-FPDS) containing 6928 annotated RGB images representing fall and no fall situations. The publicly available fall datasets are predominantly from wearable devices and camera-based sensors. Specifically, there is no publicly available audio fall dataset. Therefore, we self-collect audio data for training and evaluating the proposed method.

\section{Audio dataset}
\label{dataset}
\subsection{Data acquisition}
Our work aims to detect falls in the bathrooms, using sound as input. Since none of the publicly available datasets represent fall instances from the target environment (bathrooms), we collect our own data to train and evaluate the proposed solution for fall detection. Specifically, an audio dataset with 8 different scenarios is collected from 12 volunteers. We carefully define scenarios that represent a realistic environment, thus including both non-verbal and verbal indications of a fall event. The volunteers represent different genders, demography, and vary between 15 to 50 years old. The participants are instructed to mock all of the 8 scenarios shown in Table \ref{tab2} to collect audio data from the bathroom environments in their own homes. All of the data is collected using smartphones. Each scenario in Table \ref{tab2} represents either a ``fall'' or a ``no-fall'' class. Some of the scenarios for the two classes are expected to be difficult for the model to predict. It ensures that the proposed solution is robust to false alarms while reliably detecting falls.

We accumulate a total of 82 audio files after the data collection step. The last column of Table \ref{tab2} has the detailed breakdown of audio files in each category. It is essential to highlight that the audio files collected are of varying lengths, with the length of the longest file being 8.74 seconds. Additionally, the activity of interest (a fall event) can occur anywhere during the audio file duration.

\begin{table*}
\caption{\label{tab2}Overview of the recorded dataset.}
\centering
\begin{tabular}{p{1.00cm}p{1.25cm}p{4.75cm}p{7.5cm}}

\hline
Category ID& Activity demonstrated (Class type) & Audio recording scenario& Recorded data breakdown  \\
\hline
1 & Fall & Running water (from shower or sink or toilet) and the person says keywords demonstrating a true Fall such as ‘Help’, ‘Misty’, ‘Call 911’, ‘Help Misty’& Total number of examples: 12

Person says ‘Help’, ‘Please Help’ in a loud tone as in a panic mode: 9

Person says ‘Help’, ‘Help Misty’, ‘Call 911’ in a normal/soft tone: 3\\
\hline
2 & No Fall & Running water (from shower or sink or toilet) and the person says some words not relevant to a Fall, thus demonstrating a No Fall situation& Total number of examples: 12

Person says, ‘I am playing with water’, ‘How are you?’, ‘Could you get me a towel?’ in a normal-loudish tone: 3

Person is talking in a monotonous tone, ‘Today is a good day’, ‘Hi, Good morning’, ‘What are you doing here’, ‘Where are you Misty’, ‘Are you there. I can't hear you’, ‘I am hungry’: 7

Person humming softly: 2\\
\hline
3 & Fall & Running water (from shower or sink or toilet) and the person screams, demonstrating a true Fall& Total number of examples: 11

Person makes a very loud sound, ‘Ahhhh’ or ‘Ouch’ or ‘Oooo’ as if hurt: 10

Person makes a softer sound, ‘Ahhh’: 1 \\
\hline
4 & No Fall & Person in the bathroom performing any activity such as cleaning bathroom/taking shower and singing or talking, etc. thus demonstrating a No Fall &Total number of examples: 11

Person says ‘I am cleaning’, ‘what are you doing here’, ‘Hey Misty, please sing me something’,  in a loud tone’: 3

Person humming softly: 3

Person singing loudly: 4

Person talking softly: 1\\
\hline
5 & No Fall & Running water (from shower or sink or toilet) with no other sound interference - just plain running water & Total number of examples: 12

Plain running water: 12
\\
\hline
6 & Fall & Running water(from shower or sink or toilet) and make a loud noise of any kind demonstrating a true Fall situation (such as a ‘bang’ when hitting the floor or wall or the door) & Total number of examples: 11 

A loud sound of something dropping on the floor:9

A softer sound of something dropping on the floor: 2
\\
\hline
8 & Fall & Quiet bathroom environment with a person banging on the bathroom door, demonstrating a true Fall situation and as if asking for help & Total number of examples: 12

Person knocks at the door and says ‘Help' in a very loud tone: 2

Person knocks at the door multiple times and loudly: 8

A soft knock at the door and person says, ‘Ahh' as if hurt: 1

A sound of something dropping/knock at the door and the person makes sound as if hurt: 1
\\
\hline
9 & Fall & Quiet bathroom environment with a person saying key words, demonstrating a true Fall situation, such as ‘Help’, ‘Misty’ & Total number of examples: 11

Person says ‘Help’, ‘Please come’, ‘Could you help?’ multiple times and in a very loud tone: 6

Person says ‘Help’, ‘Call 911’, Misty Help me’ in a normal/soft tone: 3

Person says ‘Help’, ‘Call 911’, ‘Call Police’ in a tone that displays distress: 2

\\
\hline

\end{tabular}
\end{table*}

\subsection{Data augmentation}
Data augmentation is a technique to increase the size of the dataset by applying various transformations to the original data. It helps the deep learning model generalize better, especially when the original dataset size is too small. We apply several transformations to fall and no-fall audio files in the original data. The specific transformations applied are outlined in Table \ref{augment}. Some of the transformations, such as time shift, time mask, and others highlighted in the table, apply only to the no-fall categories. It is because these transformations can potentially destroy the legitimacy of the original data. For example, the time-shift transformation can potentially altogether remove a fall event from the audio file, making the augmented data for that instance useless. Specifically, we apply 14 audio transformations to each fall category and 100 transformations to each of the no-fall categories. After performing data augmentation, the dataset's size increases from 82 to 4390 data points. It includes 855 fall and 3535 no-fall data points. The Audiomentations\footnote{\url{https://github.com/iver56/audiomentations}.} Python library is used to apply all audio augmentations.

\begin{table*}
  \caption{Data augmentation}
  \label{augment}
  \begin{tabular}{ll}

    \toprule
    Transformation type&Description\\
    \midrule
     Gaussian noise & Add noise to the samples\\
     Gain & Increase/decrease the audio volume\\
     Gain transition & Gradual change in volume over a specific time span\\
     Loudness normalization & Apply a constant amount of Gain\\
     Pitch shift & Increase/Decrease the pitch\\
     Resample & Re-sample audio\\
     Time stretch& Increase/Decrease the audio speed\\
     augment1$^a$, augment2, augment13, \\augment,14, augment15 & Apply various combinations of the transformations at once\\
     $^*$Time shift & Shift audio left/right in time-axis\\
     $^*$High Pass, Peaking, Low Pass, \\ Band Pass, Band Stop,\\High Shelf, Low Shelf filters & Apply various filters individually\\
     $^*$Gaussioan SNR & Add gaussian noise to the input\\
     $^*$Reverse & Apply time inversion\\
     $^*$Clipping Distortion & Clip a random percentage of samples to distort audio\\
     $^*$Polarity Inversion & Multiply the audio samples by (-1)\\
     $^*$Tan Distortion & Apply tanh distortion \\
     $^*$Time Mask & Silence a random portion of the audio\\
     $^*$Normalize & Apply a peak normalization\\
     $^*$MP3 Compression & Lower the audio quality by compressing\\
     $^*$7-band parametric equalizer & Adjust volume of various frequency bands\\
     $^*$augment3 - augment12, \\augment16-augment60 & Apply various combinations of the transformations at once\\
  \bottomrule
  \footnotesize{$^*$ These transformations are applied to  ``no-fall'' categories only.}\\
  \footnotesize{$^a$ augment1 $=$ Gain transition, Add Gaussian noise, Pitch shift in sequence.}\\
  \footnotesize{Some transformations are applied more than once with different parameters.}\\
\end{tabular}
\end{table*}

\section{Methodology}
\label{methodology}
The augmented dataset is pre-processed and used to train and evaluate the proposed Transformer-based deep learning model. The tasks related to audio processing, such as reading raw audio files and feature extraction, are performed using librosa\footnote{\url{https://librosa.org/doc/latest/index.html}.} Python library.

\subsection{Data Pre-processing}
\label{preprocessing}
The raw audio files of varying lengths are read with a sampling rate of 16kHz, which is sufficient for audio processing \cite{berg2021keyword}, \cite{baevski2020wav2vec}, \cite{droghini2018few}. The sampling converts raw audio to a 1-Dimensional vector (a sampled version of the original audio) containing t$*$16k discrete samples, where "t" is the duration of the original raw audio. For example, if the audio file's duration is 8.74 seconds, it results in a 1-D vector of length (8.74$*$16k).

Given that the original audio files in our dataset are of varying lengths, the resulting 1-D vectors after sampling are also of unequal lengths. Therefore, we post-pad the shorter vectors with zeros to ensure that all data points in the dataset are of equal length. Finally, the dataset is split into three subsets: train, validation, and test sets, with an 80:10:10 ratio. This ratio is maintained for all the categories in Table \ref{tab2}. It implies that a percentage of all fall/no-fall categories are present in each set. After splitting, the train set contains 684 falls and 2826 no-falls; the validation set contains 84 falls and 353 no-falls; the test set contains 87 falls and 356 no-fall data points. While the train and validation sets are used to train and evaluate the model during the training phase,  the test set is reserved for the final evaluation of the trained model.

\subsection{Feature extraction}
The most commonly used audio features for deep learning models include log mel spectrograms \cite{kong2020panns}, \cite{gong2021ast} and Mel-frequency cepstrum coefficients (MFCC) \cite{berg2021keyword}, \cite{9383491} features. The MFCC features are typically used for processing speech audios, whereas the log mel spectrograms are used for a broader class of audios. These features are considered hand-crafted features as one must define various parameters\footnote{\url{https://librosa.org/doc/main/generated/librosa.feature.melspectrogram.html}.} such as hop length, window type, and a number of mel bins to extract such features. This strategy affects the quality of the extracted features and hence the model output. Some methods completely bypass manual feature extraction and directly feed the sampled audios to the model for training and evaluation \cite{baevski2020wav2vec}. In such cases, the deep learning model takes the raw input (in a sampled form), and it inherently learns features during the training phase that it deems are important for the task.

In our work, we study three different types of features. Specifically, we use log mel spectrograms, raw audios, and a novel set of features for audio, which we call ``Diff'' features. The description of each feature type is below.

\subsubsection{Segmented raw audio}
\label{raw}
The first feature type is a segmented version of the original sampled audio. Specifically, the original sampled audio waveform of length t seconds is broken into sequences of length t\_seg seconds. Each sequence of t\_seg seconds is then stacked vertically to create a 2-D input of size (N X D), where N is the total number of sequences and D is the number of discrete audio samples in each sequence. For example, if the original audio of 8.74 seconds (or 139760 discrete audio samples) long is broken into sequences of length t\_seg=0.1 sec, that results in a segmented sampled audio input of shape (87x1600). Note that a small portion of samples (in this case, 560 discrete samples or 0.035 sec worth of audio) is dropped from the original audio to ensure that the last sequence contains the same number of samples as the other sequences in the (NxD) matrix. It is acceptable as the last few samples correspond to zero padding in most of the audios in our dataset. Fig. \ref{features} (b) shows an example of segmented audio derived from an original audio signal.
\subsubsection{Diff features} 
\label{Diff}
The second type of feature explored in this paper derives from the segmented sampled audio described above. The idea behind creating this set of features is to capture the change between two consecutive audio sequences. Therefore, we subtract two successive sequences of t\_seg seconds to create a new sequence, which captures the difference between two successive sequences. It is repeated for all sequences in the segmented raw audio. It results in a new matrix of shape (N-1 x D), which we call ``Diff'' features. These features are fed into the deep learning model as input. Fig. \ref{features} (c) shows an example of a ``Diff'' feature derived from the segmented raw audio.
\subsubsection{Log mel spectrogram}
\label{logMel}
\begin{figure*}[ht]
  \centering
  \includegraphics[width=\linewidth]{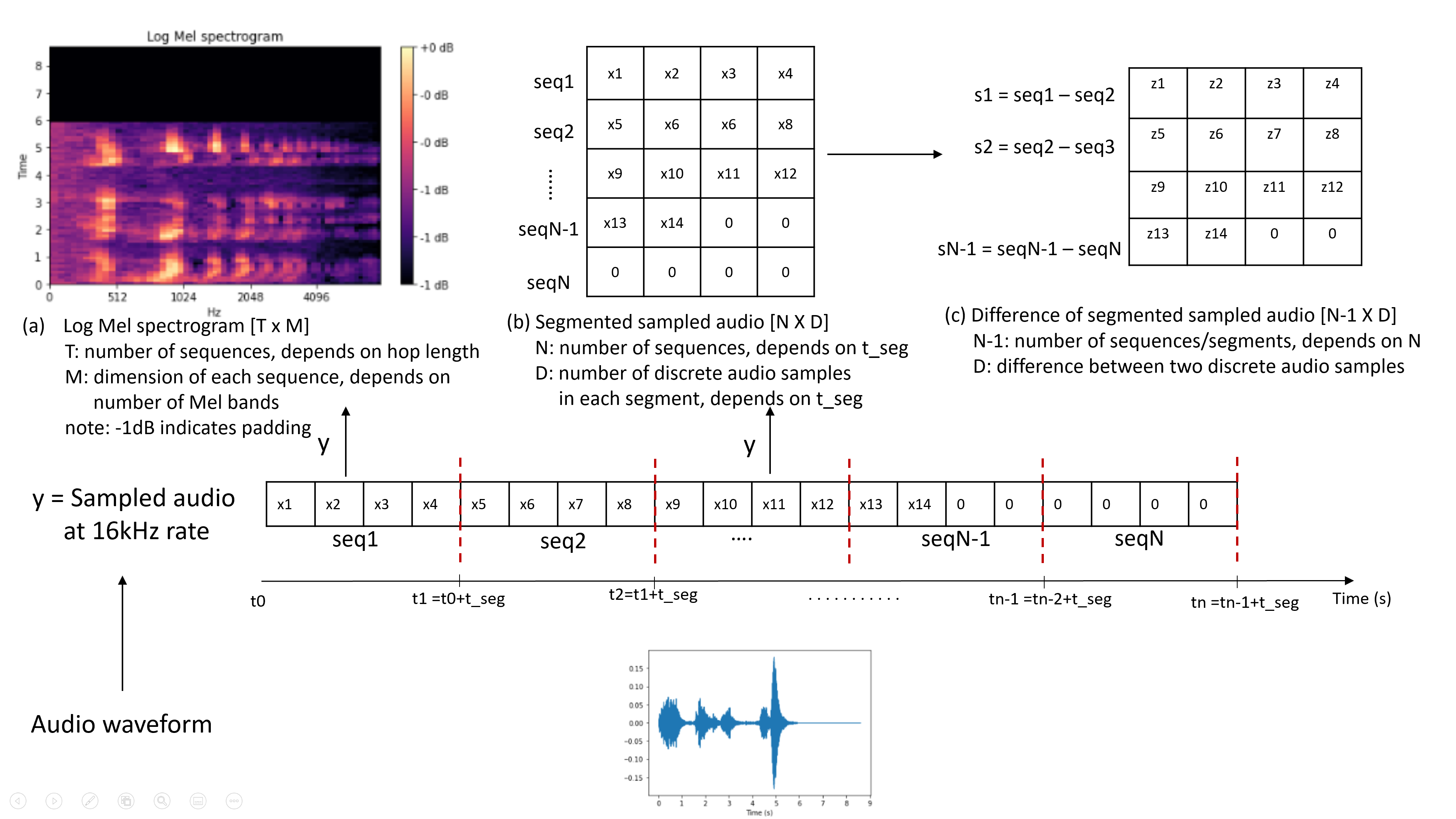}
  \caption{Three types of features (a,b,c) used for training the model.}
   \label{features}
\end{figure*}

Finally, we also explore the log mel spectrogram feature. A log mel spectrogram is attained by taking the Fourier transform (Fast Fourier Transform (FFT)/ Short Time Fourier Transform (STFT)) of the original audio signal in the time domain, then converting the resulting spectrogram to a Mel-scale, followed by applying a logarithmic operation to the Mel-scaled spectrogram. Several parameters, including the length of the STFT window, number of mel-bins, window type, and hop length (number of samples between consecutive frames), directly affect the shape and form of the mel-spectrogram and hence the performance of the deep learning model \cite{kong2020panns}. The resulting log mel spectrogram is of shape (TxM), where T is the number of frames and M is the number of mel bins. Based on \cite{kong2020panns}, \cite{gong2021ast} and experiments on our own data (Tables \ref{Diff_MelBins}, \ref{Diff_Hop_length}), we select 64 mel bins, hop length of 1600 (which is equivalent to 100ms resolution for a sampling rate of 16kHz) and ``Hann'' window type. This configuration results in a log mel spectrogram of shape (88x64) for an audio signal that is 8.74sec long. Note: The shorter hop length corresponds to higher resolution and leads to more frames (T). Similarly, more number of mel bins leads to a higher value for M. Fig. \ref{features} (a) shows an example of a log mel spectrogram derived from an original audio signal.

Next, these three features are fed independently into the deep learning model for further training and testing.

\subsection{Model Architecture}
\label{modelArchitecture}
Our model comprises of three main blocks: an input block, a feature encoder block, and an output block. While the feature encoder block used remains the same regardless of the input type fed to the model, the input and output blocks vary slightly, yielding three different overall model configurations.  
\subsubsection{Model configuration A}
\label{configA}
In this model configuration, we use segmented raw audio features from Section \ref{raw} as an input. The input size is (87x1600), where 1600 is the feature dimension. Since we are solving a classification problem, we append a CLS token at the beginning of the input to learn a global representation of the entire input \cite{devlin2018bert}. The CLS token is initialized to 1 and has the same dimension as the feature dimension of the input, which in this specific case is 1600. At this stage, the input shape is (88x1600). It is then followed by a linear projection which maps the features to a lower dimension, \textit{d\_model}. We choose \textit{d\_model} as 512 as in \cite{vaswani2017attention}. It results in input of size (88x512). Finally, we add a sinusoidal position encoding to the input since the sequence order is important. This completes the input block of the model. The input with the positional encoding is fed to the feature encoder block.

The feature encoder block of our model is composed of several encoder layers of a standard Transformer architecture \cite{vaswani2017attention}. Each Transformer encoder layer consists of Multi-Head Attention (MHA) and feed-forward sub-layers. We select the number of encoder layers (\textit{Nt}) and the number of MHA heads (\textit{H}) as 12 and 12, respectively, based on the ablation studies we performed (Tables \ref{Transformer_studies_logMels}, \ref{Transformer_studies_raw}- \ref{Transformer_studies_combined}). In addition to the input received from the input layer of the model, the feature encoder block also receives a binary input mask. The input mask tells the encoder block which input sequences to ignore. It is important because the shorter audios in the dataset have been post padded with zeros as explained in Section \ref{preprocessing}. If the input mask is not provided, the encoder will not be able to distinguish between the sequences of the input that contain the actual audio data and those that contain the zero padding. During the training phase, the feature encoder block learns the embedding for the input. The dimension of the learned embedding is (88x512). We extract the embedding of the CLS token from the output of the feature encoder block and feed it to the output block for further processing and final classification.

The output block of the model consists of a Multi-Layer Perceptron (MLP), followed by a final dense layer for classification. The MLP is a sequence of dense layers containing 265, 64, and 10 neurons. The output of the MLP feeds to the final classification layer, which consists of 2 neurons. The final classification layer outputs logits for two possible classes, namely the ``fall'' and ``no-fall'' classes. These logits are passed through the softmax activation to output the predicted final class.

\subsubsection{Model configuration B}
This model configuration uses Diff features from Section \ref{Diff} as an input. The size of the input fed to the model is (86x1600). The rest of the model architecture is the same as configuration A, described in Section \ref{configA}. Specifically, we first add a CLS token to the input, perform a linear projection, and finally feed the input plus the positional encoding to the feature encoder block. The embedding of the CLS token is extracted from the feature encoder block and passed to the output block. Within the output block, the embedding of the CLS token passes through an MLP layer and a final dense layer for classification.

\subsubsection{Model configuration C} 
The last configuration uses log mel spectrogram features, described in Section \ref{logMel}. The input fed to the model has the size (88x64). This configuration is a simpler version of the model configurations A and B. Specifically, we do not perform any linear projection in the input block and remove the MLP layer from the output block. The \textit{d\_model} is set to 64. Other necessary changes include changing the dimension of the CLS token to 64. The embedding of the CLS token extracted from the feature encoder block is passed to the final dense layer directly for classification.

The model construction, training, and evaluation are configured using the Keras API and Tensorflow 2.0 in a Google Colab environment. The hyper-parameters chosen for the training phase are per Table \ref{Hyperparameters}. Finally, the trained model is saved on a local machine for use during inference at a later time. The classification results of the trained model are discussed in Section \ref{results}.

\begin{figure*}[h!]
  \centering
  \includegraphics[width=0.8\linewidth]{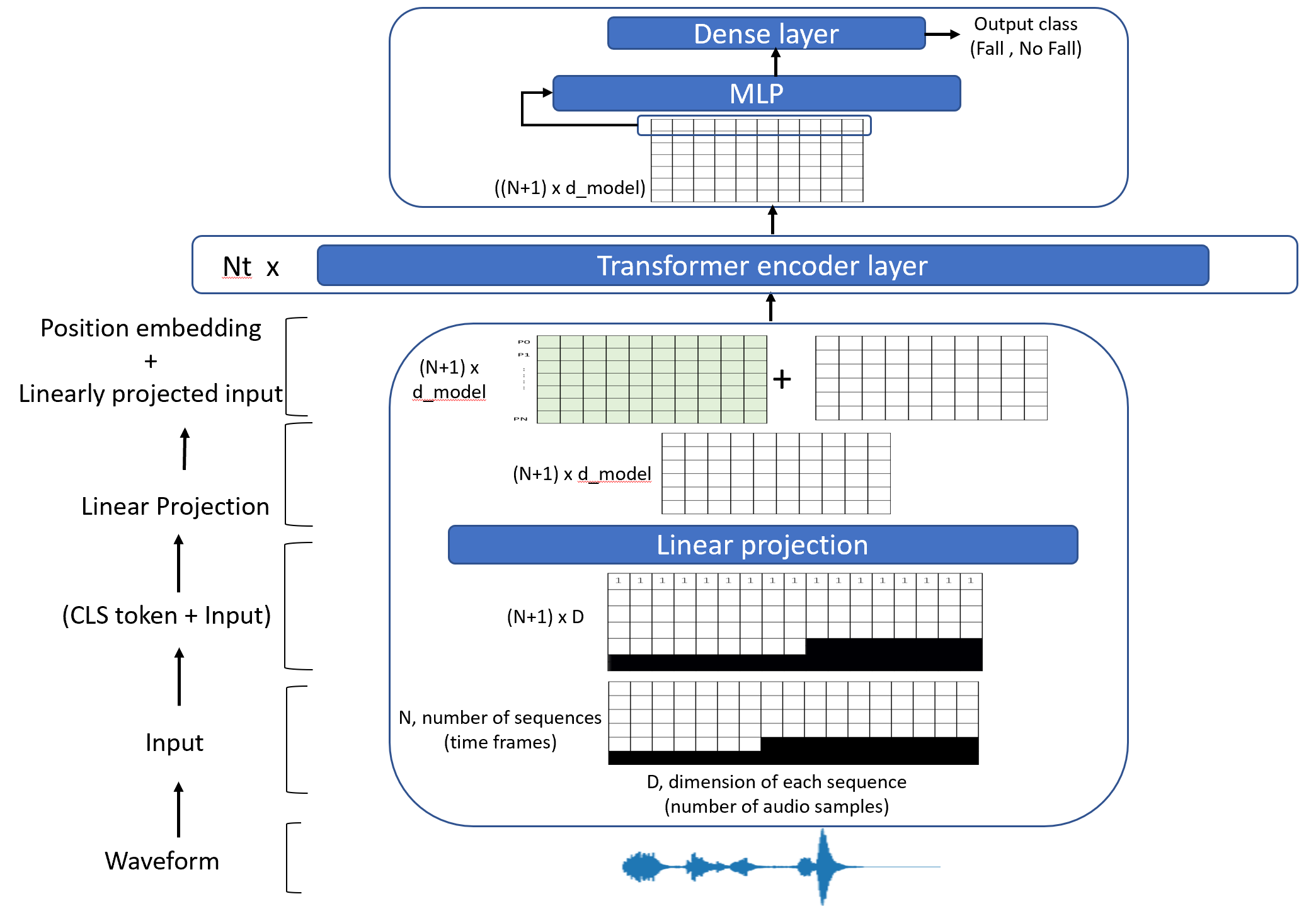}
  \caption{The Configuration A of a proposed Transformer-based architecture consists of three major blocks of the model. The raw waveform is sampled, post-padded with 0s, and then segmented to derive segmented raw audio features. The input block adds a CLS token to the input, followed by a linear projection. Then the input with its position encoding is passed to the feature encoder block. After passing the input through the feature encoder block, the embedding of the CLS token is passed through an MLP and final dense layer to make the final class prediction.}
  \Description{Overall model architecture.}
  \label{modelArch}
\end{figure*}

\begin{table}
  \begin{threeparttable}
    \caption{Hyperparameters}
     \label{Hyperparameters}
     \begin{tabular}{llllll}
        \toprule
         Hyperparameter& $^a$raw audio and Diff & $^b$log mel spectrogram\\
        \midrule
         Batch size& 20&20\\
         Epochs& 10&10\\
         Optimizer& Adam&Adam\\
         Learning rate& 0.00001&0.00001\\
         Encoder layers& 12&12\\
         Number of heads& 12&6\\
         d\_model& 512&64\\
         Feed forward& d\_model*2=1024&d\_model*2=128\\ 
         dropout rate& 0.1&0.1\\ 
        \bottomrule
     \end{tabular}
     \begin{tablenotes}
      \small
      \item [a] Hyperparameters used when using Segmented raw audio and Diff features as inputs.
      \item [b] Hyperparameters used when using log mel spectrograms as inputs.
    \end{tablenotes}
  \end{threeparttable}
\end{table}

\section{Implementation}
\label{implementation}
\begin{figure}
\centering
\includegraphics[width=1.0in]{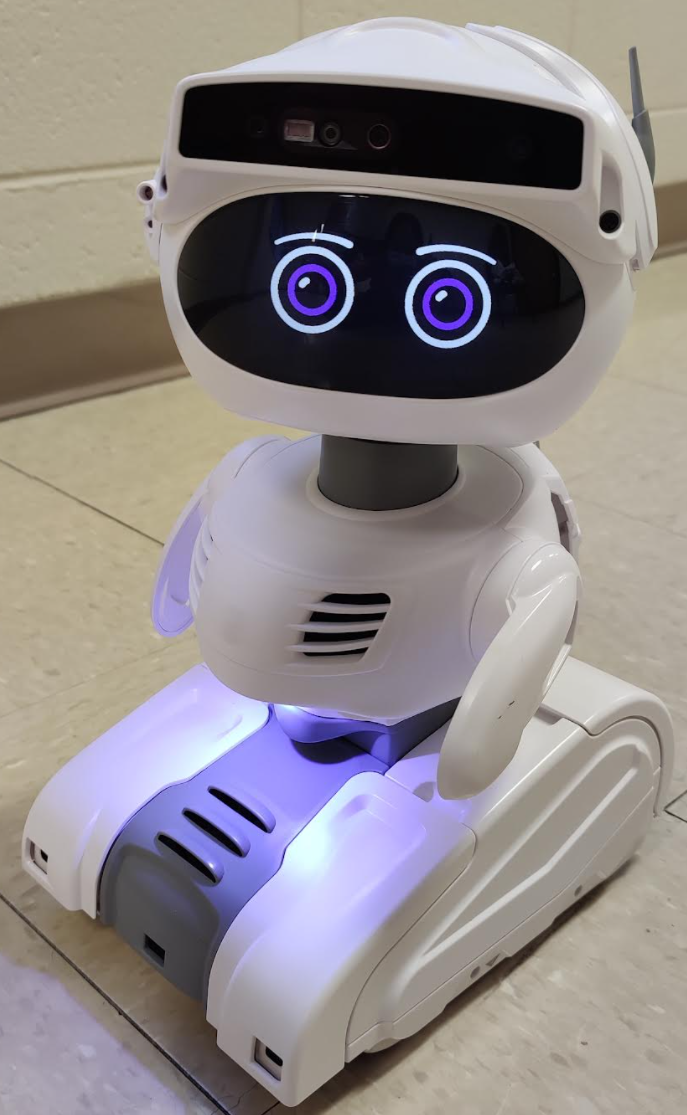}
\caption{Misty II robot.}
\label{misty}
\end{figure}

Our intended application for fall detection is to create a robot guard for an indoor home environment. Therefore, we deploy the trained model on a Misty II robot shown in Figure \ref{misty}. It is a mobile robot that is 0.35m in height, equipped with several sensors. We rely on three far-field microphones located on Misty's head for fall detection. These microphones record the ambient sound from the target environment (bathrooms) and send it to the saved model for inference. If a potential ``fall'' is detected, the robot sends a text message to a registered phone number. It serves as an alert to the caregiver. While we choose to implement the proposed method on a mobile robot as it offers an added advantage of broader sensor coverage due to its mobility, this method can be implemented on other devices as long as it has access to a microphone. 

\section{Results and Discussion}
\label{results}
This section discusses the results of the proposed model in Section \ref{modelArchitecture}. We use accuracy and F1-score as the metrics to assess our model. These metrics are commonly used in research within fall classification domain \cite{fei2022flow}, \cite{de2022bluetooth}, \cite{droghini2018few}, \cite{8891779}. We label the ``fall'' class as a positive class and the ``no-fall'' class as a negative class. The definitions of the metrics used are given in the equations below. Note that TP, TN, FP, and FN are True Positive, True Negative, False Positive, and False Negative, respectively.


\begin{equation}
  Accuracy = \frac{TP + TN}{TP+TN+FP+FN}  
\end{equation}

\begin{equation}
  F1 score = \frac{2*Precision*Recall}{Precision + Recall}
\end{equation}

\begin{equation}
  Recall = \frac{TP}{TP + FN} 
\end{equation}

\begin{equation}
  Precision = \frac{TP}{TP + FP} 
\end{equation}

\begin{figure}[h!]
\centering
\includegraphics[width=2.5in]{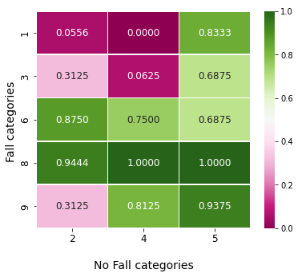}
\caption{Fall/No-Fall categories pairwise prediction results (Recall) using Diff features and weighted loss.}
\label{pairwise_analysis}
\end{figure}

Table \ref{overall_results} shows the overall results using different features and the Transformer model configurations discussed in the paper. We use a Support Vector Machine (SVM) and a Dense Neural Network (DNN) as baselines to compare our results as in \cite{droghini2018few}. Further, for the SVM, both linear and rbf kernels are evaluated. The ``Diff'' features give the best results in terms of both accuracy and F1-score, with an accuracy of 0.8673 and F1-score of 0.6234. Further, our model has a higher F1-score than the SVM and DNN, regardless of the features used. While the model outperforms both SVM and DNN, the F1-score is lower than desired.
To investigate deeper, we conduct pairwise analysis on all of the categories in Table \ref{tab2}. The results in terms of a recall value for the fall class are shown in Figure \ref{pairwise_analysis}. It is evident that the model can differentiate between most fall and no-fall categories. The instances where it struggles to give high recall is when \textbf{both} fall and no-fall categories contain human speech. It includes category pairs 1 and 2, 1 and 4, 3 and 2, 3 and 4, 9 and 2. It implies that the model cannot distinguish between a normal human speech versus a speech that contains distress calls such as calling for help. Further, the model generally has a higher recall and a lower precision. For example, it gives a recall of 0.82 and 0.65 at precision of 0.3, using diff and log mel spectrogram features, respectively. In our future work, we will consider adding human speech recognition and emotion recognition from sound to be able to handle such challenging cases.

Next, we compare the results with other state-of-the-art works in this domain. As no public audio dataset for fall detection is available for evaluation, a direct comparison with other works is not possible. However, we review recent works that use audio input to detect falls to compare our work. \cite{collado2017machine} uses several machine learning techniques to detect falls. The random forest and logistic regression yield an F1-score of 86.5\%. It is important to emphasize that the fall class in the dataset used in \cite{collado2017machine} only considers fall sounds that result from an impact on the floor when a person falls on the ground. The falls are simulated by creating a trip-over scenario and recording the sound of an impact of a person falling on the ground as a result. The sounds representing no-falls contain war sounds and normal human conversations. Further, \cite{droghini2018few} and \cite{8891779} present a more recent contribution. The authors perform data acquisition using a special floor acoustic sensor to collect sounds of various objects falling on the ground, including a human fall, book, ball, chair, and aerial microphones to record other everyday sounds such as keys, nipper, and guitar slide. The proposed method trains a Siamese Neural Network to generate embeddings used by a kNN classifier to classify a sound into fall or no-fall. The method achieves an F1-score of 93.58\%.

Based on Figure \ref{pairwise_analysis}, we believe that our method is at par with the above-mentioned state-of-the-art works if human speech data is excluded from our dataset. However, it is crucial to include human sounds as it provides an additional modality for detecting falls. Specifically, when a fall is accompanied by a scream or a verbal call for help, it would be prudent for the model to detect it in addition to a sound made by an impact on the floor. The following section conducts several experiments to identify model hyper-parameters that give the best results.

\begin{table*}[h!]
  \begin{threeparttable}
    \caption{Results on the Test set comparing different features and methods}
    \label{overall_results}
     \begin{tabular}{lllll}
        \toprule
         Method& Features & Input shape&Accuracy& F1-score \\
        \midrule
    SVM (linear kernel) & Diff & 86x1600 &0.7254 &0.2105\\
    SVM (rbf kernel) & Diff & 86x1600  & 0.8169&0.0908\\
    DNN & Diff & 86x1600 & 0.8054&0.1237 \\
    Transformer \textbf{(Ours)} & Diff & 86x1600& \textbf{0.8673}&\textbf{0.6234}\\
    \midrule
    SVM (linear kernel) & segmented raw audio & 87x1600 &0.7071 &0.1352\\
    SVM (rbf kernel) & segmented raw audio& 87x1600 &0.8169 &0.0909\\
    DNN & segmented raw audio & 87x1600 & 0.6887& 0.2765\\
    Transformer \textbf{(Ours)} & segmented raw audio & 87x64 & 0.8444&0.5405\\
    \midrule
    SVM (linear kernel) & log mel spectrograms & 88x64 &0.7620 &0.3658\\
    SVM (rbf kernel) & log mel spectrograms & 88x64 &0.7963 &0.3101\\
    DNN & log mel spectrograms & 88x64 & 0.7848&0.2166 \\
    Transformer \textbf{(Ours)} & log mels spectrograms & 88x64 &0.6796 &0.5204\\
    Transformer \textbf{(Ours)} & log mel spectrograms +segmented raw audio & 87x1664 & 0.7346&0.5704\\
        \bottomrule
     \end{tabular}
  \end{threeparttable}
\end{table*}

\subsection{Ablation studies}
     
The Transformer encoder used in this work consists of a standard encoder from the original Transformer model \cite{vaswani2017attention}. The number of encoder layers and the number of attention heads are two hyperparameters we study first. Tables \ref{Transformer_studies_logMels}, \ref{Transformer_studies_raw}, and \ref{Transformer_studies_Diff} show the model performance for different combinations of attention heads and encoder layers for all three features.

\begin{table}
  \begin{threeparttable}
    \caption{Results of different heads and encoder layers, using log mel spectrograms as an input on the Test set.}
    \label{Transformer_studies_logMels}
     \begin{tabular}{llll}
        \toprule
         Heads& Layers& Accuracy&F1-score \\
        \midrule
        1&3 & 0.7620 & 0.2877\\
        1&6 & 0.6957 & 0.3575\\
        1&12& 0.6796 & 0.5204\\
        
        3&3 & 0.7689 & 0.3764 \\
        3&6 & 0.6682 & 0.3317\\
        3&12& 0.6430 & 0.4306\\
        
        6&3 & 0.7712 &  0.3749\\
        6&6 & 0.6911 &  0.3892\\
        6&12& 0.6987 &  0.4895\\
        
        12&3& 0.6842 & 0.3610 \\
        12&6& 0.6636 &  0.3901\\
        12&12& 0.6705 & 0.5067\\
        \bottomrule
     \end{tabular}
    \begin{tablenotes}
      \small
      \item [a] Log mel spectrogram parameters: Window length=2048, Hop length=1600, Mel bins 64
      \item [b] Input shape (TxM) = 88x64
      \item [c] Transformer parameters: All other hyperparameters used are per Table \ref{Hyperparameters}
    \end{tablenotes}
  \end{threeparttable}
\end{table}

Further, for the log mel spectrograms, the number of mel bins and hop length are additional parameters affecting model performance. Ideally, a higher number of mel bins and shorter hop length should yield better results \cite{kong2020panns}, which comes at a higher computational cost. We see this trend in Table \ref{Diff_Hop_length}. However, the number of mel bins used did not increase model performance beyond the mel bin size of 64. Considering the computational cost and model performance, we select 64 mel bins and 1600 as hop length for the final model that uses log mel spectrograms as input. Next, for the segmented raw audio and Diff features, we study the segment length (t\_seg) effect on the model performance.

Table \ref{segment_lengths_Diff} shows the model performance for different segment lengths. We choose 100ms as the segment length for segmented raw audio and Diff features for the final model. 

\begin{table}
  \begin{threeparttable}
    \caption{Results of different Mel bins for log mel spectrograms on the Test set.}
    \label{Diff_MelBins}
     \begin{tabular}{llll}
        \toprule
         Mel bins& TxM& Accuracy&F1-score \\
        \midrule
    32&88x32 & 0.6476 &0.3839\\   
    64&88x64&  0.6987 &0.4895\\
    128& 88x128 & 0.5217  &0.4456\\
        \bottomrule
     \end{tabular}
    \begin{tablenotes}
      \small
      \item [a] Other log mel spectrogram parameters: Window length=2048, Hop length=1600
      \item [b] Transformer parameters: \textit{d\_model}=M, rest of the hyperparameters used are per Table \ref{Hyperparameters}
    \end{tablenotes}
  \end{threeparttable}
\end{table}

\begin{table}
  \begin{threeparttable}
    \caption{Results of different Hop length for log mel spectrograms on the Test set.}
    \label{Diff_Hop_length}
     \begin{tabular}{lllll}
        \toprule
         Hop length& Time resolution & TxM& Accuracy&F1-score \\
        \midrule
    1600&100 ms& 88x64& 0.6987 & 0.4895 \\
    1000&62.5 ms &140x64 & 0.6751 & 0.5232\\
    500&31.25 ms &280x64 & 0.7071 & 0.5428 \\
        \bottomrule
     \end{tabular}
    \begin{tablenotes}
      \small
      \item [a] Other log mel spectrogram parameters: Window length=2048, Mels = 64
      \item [b] Transformer parameters: All hyperparameters used are per Table \ref{Hyperparameters}
    \end{tablenotes}
  \end{threeparttable}
\end{table}

\begin{table}
  \begin{threeparttable}
    \caption{Results of different heads and encoder layers, using segmented raw audio as an input (t\_seg = 100ms) on the Test set.}
    \label{Transformer_studies_raw}
     \begin{tabular}{llll}
        \toprule
         Heads& Layers& Accuracy & F1-score\\
        \midrule
        1&3  &  0.8352  &0.5198 \\
        1&6  & 0.8352   &0.4626\\
        1&12  & 0.8101  &0.5146 \\
        
        3&3&   0.8009  &0.4238\\
        3&6&  0.8352   &0.5263\\
        3&12  & 0.7506  &0.3699\\
        
        6&3&  0.8444   &0.3334\\
        6&6&  0.7140    &0.4239\\
        6&12 & 0.8261   &0.4571 \\
        
        12&3  & 0.8352    &0.4857\\
        12&6 & 0.8444   &0.5405 \\
        12&12 & 0.8444   &0.5278\\
        \bottomrule
     \end{tabular}
    \begin{tablenotes}
      \small
      \item [a] Input shape (NxD) = (87x1600)
      \item [b] Transformer parameters: All other hyperparameters used are per Table \ref{Hyperparameters}
    \end{tablenotes}
  \end{threeparttable}
\end{table}

        
        
        

\begin{table}
  \begin{threeparttable}
    \caption{Results of different heads and encoder layers, using Diff features an input on the Test set.}
    \label{Transformer_studies_Diff}
     \begin{tabular}{llll}
        \toprule
         Heads& Layers& Accuracy & F1-score\\
        \midrule
        1&3 & 0.7643  & 0.3905\\
        1&6 & 0.8467  & 0.5248\\
        1&12  & 0.7849  & 0.5154 \\
        
        3&3 & 0.7941 & 0.4706\\
        3&6 & 0.8490  & 0.5479\\
        3&12 & 0.7803  & 0.3924 \\
        
        6&3 & 0.8238 & 0.4832\\
        6&6  & 0.7941  & 0.4444\\
        6&12 & 0.8146  & 0.5318\\
        
        12&3&  0.8581  & 0.5156\\
        12&6& 0.8284  & 0.5614\\
        12&12& 0.8673  &0.6234\\
        \bottomrule
     \end{tabular}
    \begin{tablenotes}
      \small
      \item [a] Input shape (N-1 x D) = (86x1600)
      \item [b] Transformer parameters: All other hyperparameters used are per Table \ref{Hyperparameters}
    \end{tablenotes}
  \end{threeparttable}
\end{table}

        
        
        

\begin{table}
  \begin{threeparttable}
    \caption{Results of different segment length, using Diff features.}
    \label{segment_lengths_Diff}
     \begin{tabular}{llll}
        \toprule
         t\_seg & NxD& Accuracy&F1-score \\
        \midrule
    50 ms& 173x800  & 0.8490 &0.6526\\   
    100 ms& 86x1600   & 0.8673 &0.6234 \\ 
    300 ms& 28x4800&0.8215   &0.2289\\
    500 ms& 16x8000& 0.9192  &0.3577\\ 
        \bottomrule
     \end{tabular}
    \begin{tablenotes}
      \small
      \item [a] Transformer parameters: All of the hyperparameters used are per Table \ref{Hyperparameters}
    \end{tablenotes}
  \end{threeparttable}
\end{table}

        
        
        

\begin{figure}[h!]
\centering
\includegraphics[width=2.5in]{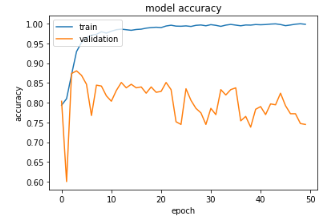}
\caption{Number of Epochs versus model accuracy on training and validation sets.}
\label{model_acc_epochs}
\end{figure}

Additionally, Figure \ref{model_acc_epochs} shows the effect of the number of epochs on both training and validation sets in terms of the model accuracy. We selected 10 epochs as the hyperparameter for the training phase based on this experiment. The model achieves an accuracy of 0.9980 on the training set after 50 epochs, whereas the validation accuracy peaks at 0.7449 after just 10 epochs. Additionally, the test accuracy after 50 epochs is 0.7620, with an F1-score of 0.4091. Finally, we study the performance of log mel spectrograms and segmented raw audios combined. Table \ref{Transformer_studies_combined} shows that the combined features do not yield any better performance than using the Diff features alone.

\begin{table}
  \begin{threeparttable}
    \caption{Results of different heads and encoder layers, using log mel spectrograms and raw features combined.}
    \label{Transformer_studies_combined}
     \begin{tabular}{llll}
        \toprule
         Heads& Layers& Accuracy & F1-score\\
        \midrule
        1&3  &0.7297 &0.4000  \\
        1&6  &0.7297 &0.5234 \\
        1&12  &0.7838&0.5512  \\
        
        3&3 &0.7838&0.5092  \\
        3&6 &0.7277 &0.4636 \\
        3&12& 0.7346&  0.5704 \\
        
        6&3  &0.7460 &0.5316 \\
        6&6 &0.8535   &0.5151  \\
        6&12   &0.8169 &0.5652  \\
        
        12&3  &0.7483 &0.5000\\
        12&6 &0.7551 &0.5485 \\
        12&12 &0.7838 &0.5569\\
        \bottomrule
     \end{tabular}
    \begin{tablenotes}
      \small
      \item [a] Input shape (NxD) = (87x1664)
      \item [b] Transformer parameters: All other hyperparameters used are per Table \ref{Hyperparameters}
    \end{tablenotes}
  \end{threeparttable}
\end{table}

\section{Conclusion and Future Work}
\label{conclusion}
In this work, we investigate ambient sound input for fall detection among the elderly. We develop and train a Transformer-based deep learning model to detect fall events from the sound input. Our method presents a non-wearable, non-intrusive and scalable solution for fall detection, especially for the bathroom environments where other existing techniques cannot be relied upon, either due to the need for wearing a device or user privacy concerns. Additionally, we implement the solution on an autonomous mobile robot that gives a broader sensing coverage due to its mobility. It is essential to highlight that the proposed solution is extensible to indoor environments beyond the bathroom with additional training. To further improve upon the results presented in the paper, we plan on collecting more variety of fall events and explore other audio features.


\bibliographystyle{ACM-Reference-Format}
\bibliography{sample-base}

\appendix

\end{document}